\documentclass[preprint,showpacs,amsfonts,aps,superscriptaddress]{revtex4} 
\usepackage{epsfig}
\usepackage{bm}

\def\be{\begin{equation}}
\def\ee{\end{equation}}
\def\bea{\begin{eqnarray}}
\def\eea{\end{eqnarray}}
\def\by{\left(\begin{array}}
\def\ey{\end{array}\right)}

\def\slash#1{\setbox0=\hbox{$#1$}#1\hskip-\wd0\dimen0=5pt\advance
       \dimen0 by-\ht0\advance\dimen0 by\dp0\lower0.5\dimen0\hbox
         to\wd0{\hss\sl/\/\hss}}

\begin{document}

\title{In-medium spectral change of $\omega$ mesons
as a probe of QCD four-quark condensate}

\author{S. Zschocke}
\affiliation{Forschungszentrum Rossendorf, PF 510119, 01314 Dresden, Germany}
\author{O.P. Pavlenko}
\affiliation{Institute for Theoretical Physics, 03143 Kiev - 143, Ukraine}
\author{B. K\"ampfer}
\affiliation{Forschungszentrum Rossendorf, PF 510119, 01314 Dresden, Germany}

\begin{abstract}
Within QCD sum rules at finite baryon density we show the 
crucial role of four-quark condensates, such as 
$\langle(\overline{q} \gamma_{\mu} \lambda^a q)^2\rangle_n$, 
for the in-medium modification 
of the $\omega$ meson spectral function. 
In particular, such a global property 
as the sign of the in-medium $\omega$ meson mass shift is found to be 
governed by a parameter 
which describes the strength of the 
density dependence of the four-quark condensate beyond mean-field 
approximation. To study self-consistently the broadening of the 
$\omega$ meson resonance we employ a hadron spectral 
function based on the 
$\omega$ meson propagator delivered by an 
effective chiral Lagrangian. Measurements of the $\omega$ meson 
spectral change in heavy-ion collisions with the HADES detector can 
reveal the yet unknown density dependence of the four-quark condensate.  
\end{abstract}

\maketitle

\section{Introduction}
The experiments with the detector system HADES \cite{lit1}
at the heavy-ion synchrotron SIS at GSI (Darmstadt)
are mainly aimed at measuring in-medium modifications of 
the light vector mesons via the ${\rm e}^{+} {\rm e}^{-}$ 
decay channel with high accuracy. 
At higher beam energies, experiments of the CERES 
collaboration \cite{lit2} at CERN SPS evidenced already hints to noticeable 
modifications of the dilepton spectrum which can be reproduced under 
the assumption of a strong melting of the $\rho$ meson in a dense, strongly 
interacting medium at temperature close to the chiral transition 
\cite{lit9,lit3,lit4}. 

The great interest in studying properties of the light mesons in a hot/dense 
nuclear medium is caused by the expectation to find further
evidences of the 
chiral symmetry restoration at finite temperature and baryon density. 
There are various theoretical indications concerning an important 
sensitivity of the meson spectral function to the partial restoration 
of the chiral symmetry in strongly interacting matter. In particular, 
at finite temperature the vector and axial-vector meson correlators 
become mixed in accordance with in-medium Weinberg sum rules 
\cite{lit5,lit6}. Such a mixing causes an increasing degeneracy between 
vector and axial-vector spectral functions which would manifest themselves
as a decrease of the $\rho$ and $A_1$ meson mass splitting, for instance. 
Similarly, the degeneracy of scalar ($\sigma$ channel) and 
pseudo-scalar ($\pi$ channel) correlators found 
in lattice QCD \cite{lit7} can lead to a considerable enhancement of the 
$\sigma$ meson spectral function at finite temperature and density 
\cite{lit8}. 

In spite of substantial efforts undertaken to understand the nature 
of vector mesons in a dense medium there is so far no unique and widely 
accepted quantitative picture of their in-medium behavior. 
The Brown and Rho conjecture \cite{lit9} on the direct interlocking
of vector meson masses and chiral quark condensate 
$\langle\overline{q} q \rangle_n$ 
supplemented by the vector manifestation of chiral symmetry in medium 
\cite{lit10,lit11} predict a strong and quantitatively the same 
decrease of the in-medium $\rho$ and $\omega$ meson masses. 

At the same time, model calculations based on various effective Lagrangians 
(cf.\ \cite{lit3}) predict rather moderate and different 
mass shifts for $\rho$ and $\omega$ mesons in a dense medium. In order 
"to match" both sets of predictions one has to go beyond simplifications made 
in the above mentioned approaches: 
The in-medium vector meson mass shift is governed 
not only by $\langle\overline{q} q\rangle_n$ but also 
by condensates of higher 
order to be calculated beyond mean-field approximation. 
Further, effective Lagrangian 
models are dealing with the scattering amplitudes in free space, so that 
the effects related to the in-medium change of the QCD condensates are hidden 
or even washed out.

The very consistent way to incorporate in-medium QCD condensates 
is trough QCD sum rules \cite{lit12,lit13}. 
As pointed out in \cite{our}, the 
in-medium mass shift of the $\rho$ and $\omega$ mesons is dominated 
by the dependence  of the four-quark condensate 
on the density. In the present letter we concentrate on the 
in-medium $\omega$ meson since the effect of the 
four-quark condensate is most pronounced for the isoscalar channel. 
Within the Borel QCD sum rule approach we go 
beyond the mean field approximation and 
use the linear density dependence of the four-quark condensate. 
We employ for the hadronic spectral function 
a constraint based on the general structure of the $\omega$ meson 
in-medium propagator 
with imaginary part of the self-energy delivered by an effective chiral 
Lagrangian. 
Our QCD sum rule evaluations show that 
the in-medium change of the four-quark condensate plays indeed
a crucial rule for modifications of the $\omega$ spectral function. 
In particular, the sign of the $\omega$ meson mass shift 
is changed by variation of a parameter
which describes the strength 
of the density dependence of the four-quark condensate.
Since the difference of the 
vector and axial vector correlators is proportional to the four-quark 
condensate, the sign of the $\omega$ 
meson mass shift, measured via the ${\rm e}^{+} {\rm e}^{-}$ channel,
can serve as a tool for determining how fast the strongly 
interacting matter approaches the chiral symmetry restoration with 
increasing density.

\section{QCD sum rule equations}

Within QCD sum rules the in-medium $\omega$ meson is considered 
as a resonance in the current-current correlation function
$
\Pi_{\mu \nu} (q , n) = i \int d^4 x \;{\rm e}^{i q x} 
\langle {\cal T} \; {\rm J}_{\mu} (x)\; {\rm J}_{\nu} (0)\rangle_n\;,
$
where $q_{\mu}=(E, {\bf q})$ is the $\omega$ meson four-momentum, 
${\cal T}$ denotes the time ordered product of the $\omega$ meson current 
operators ${\rm J}_{\mu} (x) {\rm J}_{\nu} (0)$,
and $\langle \cdots \rangle_n$ stands for the expectation value 
in the medium. In what follows,
we focus on the ground state of the baryonic matter approximated 
by a Fermi gas with nucleon density $n$. In terms of quark field operators, 
the $\omega$ meson current is given by 
${\rm J}_{\mu} (x)= \frac{1}{2} (\overline{\rm u} \gamma_{\mu} {\rm u} 
+ \overline{\rm d} \gamma_{\mu} {\rm d})$. 
At zero momentum, ${\bf q}=0$, in the rest frame of the matter 
the correlator can be reduced to
$\frac{1}{3} \Pi_{\mu}^{\mu} (q^2, n) = \Pi(q^2, n)$ for $q^2<0$. 
The correlator 
$\Pi(q^2, n)$ satisfies the twice subtracted dispersion relation, which  
can be written with $Q^2 \equiv -q^2 = -E^2$ as
\bea
\frac{\Pi (Q^2)}{Q^2} = \frac{\Pi (0)}{Q^2} - \Pi^{'} (0) - Q^2 
\int_0^{\infty} \; ds \frac{R(s)}{s (s + Q^2)}\;,
\label{eq_10}
\eea
with $\Pi (0) = \Pi (q^2=0, n)$ and 
$\Pi^{'} (0) = (d \Pi (q^2) / d q^2)|_{q^2=0}$ 
as subtraction constants and 
$R(s) = - {\rm Im} \Pi (s, n) /( \pi s)$.

As usual in QCD sum rules (QSR)
\cite{lit12,lit13}, for large values of $Q^2$ one can evaluate 
the l.h.s. of eq.~(\ref{eq_10}) by the operator product expansion (OPE) 
$\Pi(Q^2) / Q^2 = - c_0 \; {\rm ln}(Q^2) + \sum_{i=1}^{\infty}
c_i / Q^{2i} $,
where the coefficients $c_i$ include the well known Wilson 
coefficients and the expectation values of the corresponding products of quark 
and gluon field operators, i.e. condensates.

Performing a Borel transformation of the dispersion relation 
eq.~(\ref{eq_10}) with appropriate mass parameter $M^2$ 
and taking into account the OPE one gets the basic QSR equation 
\bea
\Pi(0) + \int_0^{\infty} d s \,R(s)\, {\rm e}^{-s/M^2} = 
M^2 c_0 + \sum\limits_{i=1}^{\infty} \frac{c_i}{(i-1)! M^{2 (i-1)}}\,.
\label{eq_20}
\eea
The general structure of the coefficients $c_i$ up to $i=3$ is given, for 
instance, in \cite{lit15}. In order to calculate the density dependence of the 
condensates entering the coefficients $c_{0 \cdots 3}$ we employ the standard 
linear density approximation, which is valid for not too large
density.
This gives for the chiral quark condensate 
$\langle \overline{q} q\rangle_n = \langle \overline{q} q\rangle_0 
+ \frac{\sigma_N}{2 m_q} n $,
where both the light quark masses and their condensates are taken to be the 
same, i.e., $m_q = m_u = m_d = 7$ MeV and 
$\langle \overline{q} q \rangle = \langle \overline{u} u \rangle = 
\langle \overline{d} d \rangle$ with 
$\langle \overline{q} q \rangle_0 = - (245 {\rm MeV})^3$.  
The nucleon sigma term is $\sigma_N = 45$ MeV.

The gluon condensate is obtained as usual employing the QCD trace anomaly
$
\langle\frac{\alpha_s}{\pi} {\rm G^2}\rangle_n = 
\langle\frac{\alpha_s}{\pi} {\rm G^2}\rangle_0 - \frac{8}{9} M_N^0 \;n,$
where $\alpha_s = 0.38$ is the QCD coupling constant and 
$M_N^0 = 770$ MeV is 
the nucleon mass in the chiral limit.
The vacuum gluon condensate is
$\langle \frac{\alpha_s}{\pi} {\rm G^2}\rangle_0 = (0.33 \, {\rm GeV})^3$.

The coefficient $c_3$ in eq.~(\ref{eq_20}) contains also the 
four-quark condensates 
$\langle(\overline{q}\gamma_{\mu}\lambda^{a} q)^2\rangle_n$
and 
$\langle(\overline{q}\gamma_{\mu}\gamma^5 \lambda^{a} q)^2\rangle_n$.
The standard approach to estimate their density dependence consists in
the mean-field approximation. 
Within such an approximation the four-quark condensates 
are proportional to $\langle\overline{q} q\rangle_n^2$ 
and their density dependence is actually 
governed by the square of the chiral quark condensate.
Keeping in mind the important role of the four-quark condensate 
for the in-medium 
modifications of the $\omega$ meson \cite{our} 
we go beyond the above mean-field 
approximation and employ the following parameterization
\bea
\langle(\overline{q} \gamma_{\mu}\gamma^5 \lambda^a q)^2\rangle_n = 
- \langle(\overline{q} \gamma_{\mu} \lambda^a q)^2\rangle_n
= \frac{16}{9} \langle\overline{q} q\rangle_0^2 \;\kappa_0 \;
\left[1+\frac{\kappa_N}{\kappa_0}\frac{\sigma_N}{m_q 
\langle \overline{q} q\rangle_0}\;n\right]\;.
\label{eq_35}
\eea
In vacuum, $n=0$, the parameter $\kappa_0$ 
reflects a deviation from the vacuum saturation assumption.
The case
$\kappa_0=1$ corresponds obviously to the exact vacuum saturation as
used, for instance, in \cite{lit15}. 
To control the deviation of the in-medium four-quark condensate 
from the mean-field approximation we introduce the parameter $\kappa_N$. 
The limit $\kappa_N = \kappa_0$ recovers the mean-field approximation, 
while the case $\kappa_N>\kappa_0$ ($\kappa_N<\kappa_0$) is 
related to a stronger (weaker) 
density dependence compared to the mean-field approximation. 
Below we vary the parameter $\kappa_N$ to estimate 
the contribution of the four-quark 
condensates to the QSR with respect to the main trends 
of the in-medium modification 
of the $\omega$ meson spectral function.

Using the above condensates and usual Wilson coefficients 
we get the coefficients $c_{0 \cdots 3}$ as 
\bea
c_0 &=& \frac{1}{8 \pi^2} \left(1 + \frac{\alpha_s}{\pi}\right),\; \quad
c_1 = - \frac{3 m_q^2}{4 \pi^2}, \;\nonumber\\
c_2 &=& m_q \langle\overline{q} q\rangle_0 + \frac{\sigma_N}{2} \;n + 
\frac{1}{24} \left[\langle\frac{\alpha_s}{\pi} G^2 \rangle_0 
- \frac{8}{9} M_N^0 \;n\right] 
+ \frac{1}{4} A_2 M_N \;n,\nonumber\\
c_3 &=& - \frac{112}{81} \pi \;\alpha_s \;\kappa_0\; 
\langle\overline{q} q\rangle_0^{2} 
\left[1+\frac{\kappa_N}{\kappa_0}\frac{\sigma_N}{m_q 
\langle \overline{q} q\rangle_0}\;n\right]
- \frac{5}{12} A_4 M_N^3 \;n\;.
\label{eq_40}
\eea
The last terms in $c_{2,3}$ correspond to the derivative condensates 
from nonscalar operators as a consequence of the breaking of Lorentz 
invariance in the medium. These condensates are proportional to the 
moments $A_i=2\int\limits_0^1 d x \;x^{i-1} \left[q_N (x , \mu^2) + 
\overline{q}_N (x, \mu^2) \right]$ 
of quark and antiquark distributions $q_N, \bar q_N$ inside 
the nucleon at a scale $\mu^2 = 1 \, {\rm GeV}^2$ 
(see for details \cite{lit13}). Our choice of the 
moments $A_2$ and $A_4$ is 1.02 and 0.12, respectively. 

To model the hadronic side of the QSR eq.~(\ref{eq_20}) we make the standard 
separation of the $\omega$ meson spectral density into resonance part 
and continuum contribution by means of the threshold parameter $s_0$:
\bea
R(s, n)= F \;\frac{S(s,n)}{s} \;\Theta(s_0-s) + c_0\; \Theta (s-s_0)\;,
\label{eq_45}
\eea
where $S(s,n)$ stands for the resonance peak in the spectral function; 
the normalization $F$ is 
unimportant for the following consideration. 
In the majority of the QCD sum rule evaluations, 
the zero-width approximation \cite{lit13} or some 
schematic parameterization of $S$  \cite{lit16} 
are employed. 
In contrast to this, we use here a more 
realistic ansatz for the resonance spectral density $S$ based 
on the general structure 
of the in-medium vector meson propagator in the vicinity to the pole mass, 
\bea
S(s, n) = - \frac{{\rm Im} \Sigma (s,n)}{(s - m_{\omega}^2 (n))^2 + 
({\rm Im} \Sigma(s,n))^2}
\label{eq_50}
\eea
with ${\rm Im} \Sigma(s,n)$ as imaginary part of the 
in-medium $\omega$ meson self-energy and 
$m_{\omega}(n)$ as its physical mass. In eq.~(\ref{eq_50}), 
the real part of the self--energy is absorbed in 
$m_{\omega} (n)$, which is determined by
the QCD sum rule eq.~(\ref{eq_20}). As a result (see below), 
the in-medium change of the QCD condensates causes 
global modifications of the $\omega$ meson spectral function, in 
addition to the collision broadening. 
(An analogous approach was used in \cite{lit17}.)

\begin{figure}[th]
\epsfig{file=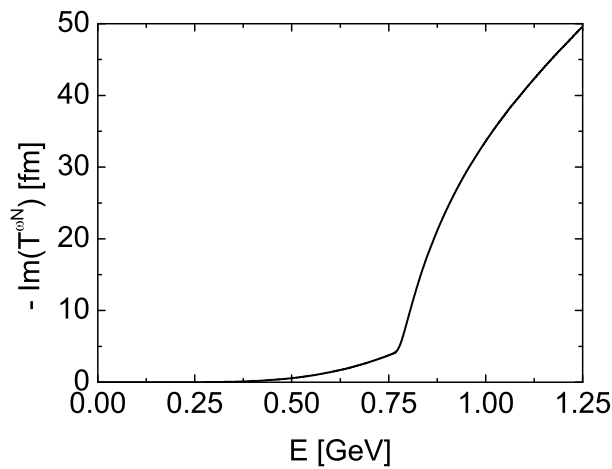,width=9cm}
\caption{Imaginary part of the off-shell $\omega$N  forward
scattering amplitude.}
\label{fig1}
\end{figure}

Within the linear density approximation the $\omega$ meson self-energy 
is given by 
\bea
\Sigma (E,n) = \Sigma^{\rm vac} (E) - n\;T^{\omega N} (E)\;,
\label{eq_55}
\eea
where $E = \sqrt{s}$ is the $\omega$ meson energy, $\Sigma^{\rm vac} (E) 
= \Sigma(E,n=0)$ and $T^{\omega N} (E)$ is the off-shell forward 
$\omega$-nucleon scattering amplitude in free space. Note, that the 
three-momentum of 
$\omega$ mesons in eq.~(\ref{eq_55}) is still zero, and nucleons are 
also assumed to be at rest. We take
the needed ${\rm Im} T^{\omega N} (E)$ from an effective Lagrangian 
\cite{lit14}, which is based on vector meson dominance and chiral 
SU(3) dynamics. 
The dominant contribution to ${\rm Im} T^{\omega N} (E)$ in 
the region $E \stackrel{<}{\sim} 1$ GeV comes from the processes 
$\omega N \rightarrow \pi N$ 
(at $E < 0.6$ GeV) and $\omega N\rightarrow \rho N\rightarrow \pi \pi N$ 
(at $E > 0.6$ GeV), see \cite{lit14} for details. 
The process $\omega N \rightarrow \pi N$ is at least partially under 
experimental control, while the contribution from the reaction 
$\omega N\rightarrow \rho N\rightarrow \pi \pi N$ appears to be 
sensitive to the poorly known meson-baryon 
form factors. However, this uncertainty does not
spoil the results below. For definiteness,
in Fig. \ref{fig1} we plot ${\rm Im} T^{\omega N} (E)$ 
employed in our QSR evaluations. 
To simplify the calculations we also take 
${\rm Im} \Sigma^{\rm vac}(E) \approx 
{\rm Im} \Sigma^{\rm vac} (E=m_{\omega}^{\rm vac}) 
= - m_{\omega}^{\rm vac} \,\Gamma_{\omega}^{\rm vac}$, 
where $m_{\omega}^{\rm vac}$ and $\Gamma_{\omega}^{\rm vac}$ 
are the vacuum $\omega$ meson mass and decay width, respectively.

Following \cite{lit14} we use for the subtraction constant 
$\Pi(0) = 9 n / (4 M_N)$ which is actually the Thomson limit of the 
$\omega N$ 
scattering process, but also coincides with the Landau damping term elaborated 
in \cite{lit18} for the hadronic spectral function entering the dispersion 
relation 
without subtractions. For details about the relation of subtraction constants 
and the Landau damping we refer the interested reader
to \cite{lit19}.

\section{Results of QCD sum rule evaluation}

Taking the ratio of eq.~(\ref{eq_20}) to its derivative with respect to 
$M^2$, and using eq.~(\ref{eq_45}) one gets
\bea
\frac{\int_0^{s_0} ds \; S(s,n)\;{\rm e}^{-s/M^2}}
{\int_0^{s_0} ds \; S(s,n) s^{-1} \; {\rm e}^{-s/M^2}} 
= 
\frac{c_0\,M^2\,[1-\left(1 + s_0 M^{-2} \right) {\rm e}^{-s_0/M^2}] 
- c_2 M^{-2} - c_3 M^{-4}}{c_0\,\left(1-{\rm e}^{-s_0/M^2}\right) 
+ c_1 M^{-2} + c_2 M^{-4} + \frac12 c_3 M^{-6} - 
\frac{9\,n}{4 M_N} M^{-2}}
\label{eq_60}
\eea
with the coefficients $c_{0 \cdots 3}$ from eq.~(\ref{eq_40}) and the 
spectral function $S(s,n)$ from eqs.~(\ref{eq_50}) and (\ref{eq_55}).
One has to solve eq.~(\ref{eq_60}) to find the mass parameter 
$m_{\omega} (n, M^2,s_0)$. At a given density $n$,
the continuum threshold $s_0$ 
is determined by requiring maximum flatness of $m_{\rm \omega} (M^2)$ within 
the Borel window $M_{\rm min} \cdots M_{\rm max}$. 
The minimum Borel mass $M_{\rm min}$ is obtained 
such that the terms of order ${\cal O} (M^{-6})$
on the OPE side contribute not more than 10\% \cite{lit16,lit20}. 
According to our experience \cite{our},
$m_{\omega} (M^2)$ is not very sensitive to 
variations of $M_{\rm max}$.
We therefore fix the maximum Borel mass by 
$M_{\rm max}^2 = 2\; {\rm GeV}^2$.
To get the $\omega$ meson mass $m_{\omega}$ we average 
finally $m_{\omega} (M^2)$ within the Borel mass window.

The value of $\kappa_0$ in eq.~(\ref{eq_35}) is related 
to such a choice of the chiral condensate $\langle\overline{q}
q\rangle_0$
which adjusts the vacuum 
$\omega$ meson mass to $m_{\omega} (n=0)=782$ MeV
resulting in $\kappa_0=3$. The ratio   
$\kappa_N/\kappa_0$ in the parameterization (\ref{eq_35}) is restricted by 
the conditions $q_4= \langle(\overline{q} \gamma_{\mu} \lambda^a q)^2\rangle_n
\rightarrow 0$ 
with increasing density and $q_4\le 0$, so that one gets 
$0 \le \kappa_N \stackrel{<}{\sim} 4$ as numerical limits. 

\begin{figure}[bh]
\epsfig{file=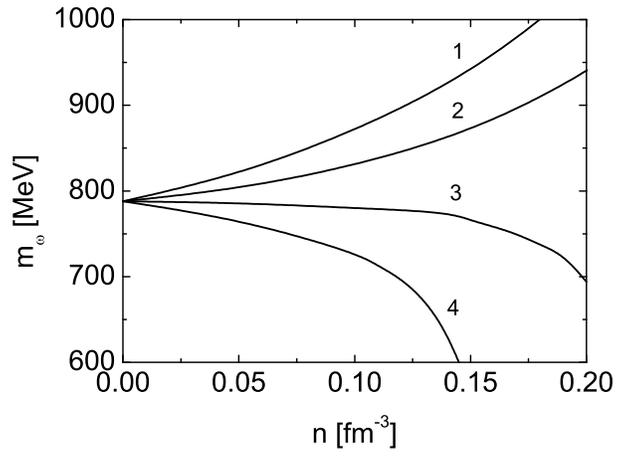,width=9cm}
\caption{Density dependence of the $\omega$ meson mass
for various values of the parameter $\kappa_N$.}
\label{fig2}
\end{figure}

The results of our QSR evaluations for $m_{\omega} (n)$ 
for $\kappa_N = 1 \cdots 4$ are exhibited in Fig.~\ref{fig2}. 
The in-medium mass of the 
$\omega$ meson changes even 
qualitatively under variation of the parameter $\kappa_N$: for 
$\kappa_N < 2.7$, $m_{\omega}$ increases with density, 
while for $\kappa_N > 2.7$ it drops.
From the above considerations one can 
conclude that the sign of the in-medium mass shift is directly related to 
the density dependence of the four-quark condensate.

\begin{figure}[th]
~\center
\epsfig{file=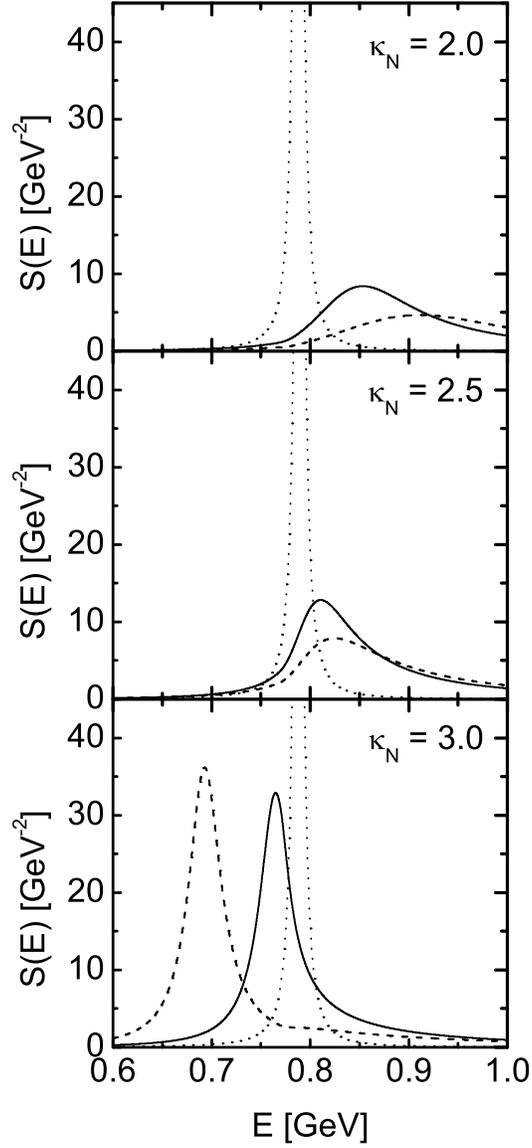,width=9cm}
\caption{$\omega$ meson spectral function for 
$\kappa_N = 2$, 2.5, 3.
Solid curves correspond to normal nuclear density 
($n=n_0=0.15 {\rm fm}^{-3}$),
while dotted and dashed curves are for vacuum and 
$n=0.2 {\rm fm}^{-3}$, respectively.}
\label{fig3}
\end{figure}

In Fig.~\ref{fig3} we display the in-medium change of the $\omega$ meson 
spectral function $S(E)$, given by eq.~(\ref{eq_50}) and calculated with 
density dependent mass $m_{\omega}(n)$. 
The in-medium spectral change is still seen 
to be dominated by the density dependence of the four-quark condensate. 
The dependence of the peak position as a function of the density $n$
and the four-quark parameter $\kappa_N$ is the same as for 
$m_\omega$. One can also observe that 
for the positive mass shift (say, for $\kappa_N=2$) 
the width increases with density, 
while for a negative mass shift (say, for $\kappa_N=3$) 
it appears to be approximately 
constant. In both cases we obtain a considerable "melting" of the in-medium 
$\omega$ meson: at $n=n_0$ the height of the resonance peak drops more than 
by a factor of 5.
For $\kappa_N = 2.5$ the shift of the peak is small; also here
the peak is broadened. 

Because of the strong ''melting'' of the in-medium $\omega$ meson, an 
identification of its spectral change in matter will be an experimental 
challenge. Moreover, in heavy-ion collisions one can expect also an 
additional broadening of the signal due to the collective expansion of the 
matter \cite{lit21}. Nevertheless, the high precision 
measurements planned with 
HADES give a chance to observe at least an ''in-medium shoulder'' which 
supplements the vacuum peak. Whether such a shoulder will occur at the right 
or left hand side of the vacuum peak is directly governed by the strength 
of the density dependence of the four-quark condensate.

\section{Summary}

In summary we have found that, within the Borel QCD sum rule approach,
the in-medium 
spectral change of the $\omega$ meson is dominated by the density dependence 
of the four-quark condensate. We go beyond the standard mean-field 
approximation and vary with the parameter $\kappa_N$ the strength of the 
four-quark condensate dependence on the density. 
The sign of the $\omega$ meson 
in-medium mass shift and the resonance peak position are shown to be 
governed by the value of $\kappa_N$. 
We find a strong ''melting effect'' of the in-medium $\omega$ meson resonance.
Only for positive mass shift we observe a considerable 
broadening of the in-medium $\omega$ meson spectral function.

The in-medium $\omega$ meson spectral change, 
in particular the sign of the mass shift, 
to be looked for
via the ${\rm e}^{+}\, {\rm e}^{-}$ channel with the 
HADES detector in heavy-ion collisions, 
can give an important information on the yet unknown 
density dependence of the four-quark condensate and consequently on the 
chiral symmetry restoration in a dense nuclear medium. \\[3mm]
\noindent{\em Acknowledgments:}
We thank E. G. Drukarev, R. Hofmann, V. I. Zakharov and G. M. Ziovjev for 
useful discussions. O. P. P. acknowledges the warm hospitality of the nuclear 
theory group in the Research Center Rossendorf. This work is supported 
by BMBF 06DR921, STCU 15a, CERN-INTAS 2000-349, NATO-2000-PST CLG 977 482.


\end{document}